# London in Lockdown: Mobility in the Pandemic City

Michael Batty[±], Roberto Murcio, Iacopo Iacopini,
Maarten Vanhoof, and Richard Milton

Centre for Advanced Spatial Analysis (CASA)
University College London, 90 Tottenham Court Road, London W1T 4TJ

November 13, 2020

## Abstract

This chapter looks at the spatial distribution and mobility patterns of essential and non-essential workers before and during the COVID-19 pandemic in London, and compares them to the rest of the UK. In the 3-month lockdown that started on 23 March 2020, 20% of the workforce was deemed to be pursuing essential jobs. The other 80% were either furloughed which meant being supported by the government to not work, or working from home. Based on travel journey data between zones (983 zones in London; 8,436 zones in England, Wales and Scotland), trips were decomposed into essential and non-essential trips. Despite some big regional differences within the UK, we find that essential workers have much the same spatial patterning as non-essential for all occupational groups containing essential and non-essential workers. Also, the amount of travel time saved by working from home during the Pandemic is roughly the same proportion – 80% – as the separation between essential and non-essential workers. Further, the loss of travel, reduction in workers, reductions in retail spending as well as increases in use of parks are examined in different London boroughs using Google Mobility Reports which give us a clear picture of what has happened over the last 6 months since the first Lockdown. These reports also now imply that a second wave of infection is beginning.

## Acknowledgements

The Future Cities Catapult funded the early stage of this research (2016-2018) and the data was assembled for the **QUANT** model which is now supported by The Alan Turing Institute under **QUANT2**–Contract–CID–3815811. Partial support has also come from the UK Regions Digital Research Facility (UK RDRF) EP/M023583/1.

[±] Correspondence this author, m.batty@ucl.ac.uk Tel 44-(0)7768-423-656; t@jmichaelbatty
http://www.spatialcomplexity.info/

## *The 2020 Pandemic in Britain*

On 23 March 2020, Britain locked down to protect its population against what by then was widely recognised as a global Pandemic. Its population watched in horror the reports from Northern Italy of rising deaths and a health system that was simply overwhelmed with serious cases requiring intensive care. Spain was not far behind while the rest of Europe was catching up fast. COVID-19 had overwhelmed the city of Wuhan where the virus was first detected in a wildlife market in early November 2019 but the serious nature of the disease was not appreciated until it was literally on our doorstep. By March, it was clear that the disease was particularly serious for older age groups whose mortality rates for those above 80 who were admitted to hospital were close to 50 percent. The Lockdown introduced in late March was designed to stop the spread of the disease, using the time honoured method of keeping people apart until the disease could be contained which was generally assumed to have occurred when the so-called R number – its rate of change – dropped below 1. Although the Lockdown was reviewed every three weeks, in fact it lasted some three months to mid-June when it became apparent that the virus had been contained. It was then deemed 'safe' to open up parts of the economy again to social interactions, notwithstanding fairly strict measures of social distancing that mandated people to keep 2 metres apart from one another, wear masks to avoid spreading the virus through respiratory means, and to wash hands frequently to remove any traces of the virus that might have been picked up from surfaces.

The restrictions imposed by the Lockdown were designed to protect the UK National Health Service which to some extent, is the most revered public service in Britain, the only function of government to have survived the dismantling of the Welfare State that proceeded apace as Britain emerged from its industrial past. The key elements of the Lockdown involved staying at home to stop the disease spreading between households. To this end, the only exceptions were shopping for basic necessities, one form of exercise a day, medical needs, care for the vulnerable, and travelling to or from work to carry out essential services. These mandates meant no household mixing, no meeting friends or family members living in separate homes, no gatherings of more than two people in public, and no social or sporting events. Schools and churches were closed. These were Draconian measures by peacetime standards and essentially put the economy and social life into cold storage. The summer months saw a gradual opening up of the economy but in a dramatically constrained way. However by the late summer in September, it was clear that a second wave could be detected through a rapid increase in testing for the virus (Murcio, 2020). Hospital admissions began to rise and at the time of writing (end October 2020) half the country is now back in some form of Lockdown, a little less severe perhaps than the original with some hospitality, schools, shops and work still open, but with strong advice to continue working from home wherever possible and with no household mixing in the most infectious hotspots. With winter approaching, the predictions are that although the shows no signs of loosening its grip and becoming less virulent, there are vaccines on the horizon whilst our medical knowledge of how to combat the disease with pharmaceutical inventions has increased. The hope is that although the number of cases may well outstrip the peak earlier in the year, the overall impact will be less severe. However assuming a vaccine becomes available by early 2021, it will take a Herculean effort to mass vaccinate an entire population.

If you lockdown an economy in the way many governments have to combat the spread of this disease, the impact on where people work, live, and entertain is dramatic. The effect of the Pandemic has and continues to have largely destroyed our quest to travel using public or group transport. Although our focus here is not on the economic impact, there are obvious changes to the locations which we traditionally visit or frequent where we engage in the routine activities of working, shopping, educating ourselves, socialising and so on, all activities that usually require us to gather together in groups of all sizes and at all scales. The mandates of social distancing operate at the most local level but these translate themselves into how we might travel more globally throughout the metropolitan area, regionally, nationally and internationally for we need to observe the mandates for keeping apart when we use public transport of any kind (Batty, 2020).

In this chapter, we will illustrate the impact of COVID-19 on changes in mobility in both the UK and in a world city where changes in where and how we travel have been dramatic. London has one of the biggest central areas of any city worldwide, diversified into financial, retail, and government functions in several distinct cores all of which have been emptied of workers since the hit. In the financial quarter – the 'square mile' or the City – which is the traditional heart of the metropolis, half a million people usually work largely in financial and legal services and cognate activities but most have been absent, working from home, for the last 6 months since the Pandemic began. In the inner part of the metropolis, the Greater London Authority (GLA) area, home to a population of some 8 million, about 40 percent of all those travelling use public transport and these services have been operating at little more than 30 percent capacity. It is estimated that 16 percent of the 5 million workers in this GLA area were classed as 'key' or 'essential workers, with permission to work during the first Lockdown. There has been an exodus of population to the outer suburbs and to the countryside, just as happened in the last great plague that hit London in 1665 when Parliament moved to Oxford. Retail activity within the City has all but ceased and many outlets have closed, reportedly more than 1,000 in the GLA area with little sign that these closures are anywhere near complete.

In some respects, a full analysis of a city under a Pandemic is an impossible task until the Pandemic ends because the restrictions on normal life are continually changing. Here, however, we will focus on mobility examining the extent to which the patterns of movement we have already noted have been disrupted by the need to social distance on all spatial scales. We will begin by examining the location patterns of essential and non-essential workers in London which are defined in terms of occupations and which are assumed to separate workers who have remained at their place of work during the Pandemic from those who are either furloughed and supported by the government (to not work) or are working from home. We will extend this analysis to how different types of workers travel between home and work under the Pandemic.

Although our analysis is largely inconclusive with respect to defining distinct differences between essential and non-essential workers in terms of their geospatial attributes, we then follow up this analysis by examining a more detailed picture of changes in travel patterns using Google's Mobility Reports. These reveal significant differences between movements associated with several physical activities ranging from transit to the use of parks and using this type of data, we can easily demonstrate how mobility varies systematically across the metropolis illustrating quite profound differences between the city core, the inner area and the outer suburbs. In fact a full analysis of all this data for cities and regions in the UK is still to be attempted but the particular case of London does provide a focus for informed speculation about how the Pandemic might end with respect to possible changes in mobility. From this data too, because it relates to how visits to different activities vary throughout the Pandemic starting in mid-February 2020 providing data each day until 18 October (the time of writing), we can see detailed change in the time series and in this way, detect the rise of the second wave and the difficulties of bringing the economy back until a vaccine is in sight. Last but not least, we will use this particular analysis to hint at ways in which London might transition to a new normal different but similar to the old normal once the Pandemic ends.

*Defining Essential Workers*

When the government locked the country down, it first defined a group of 'key' or 'essential' workers whose endeavours were required to keep the country running. It produced a list of such workers defined in terms of the proportion of the numbers of persons in the 9 occupational classes defined by the Office of National Statistics and used in the Population Census. By applying the proportions to the numbers in each of the occupational classes, it is possible to derive the numbers of essential workers in each class and it is possible to do this at the level of the standard regions in the UK of which there are 12. This is the finest level of granularity we have for occupational classes at the level of the resident population which we need to work with because we need to disaggregate our trip patterns for the journey to work by occupational class so that we can generate the distribution of such classes at the workplace end of the trip. In short, our basic data involves the flows of workers from their place of residence to their place of work, data collected by the UK Population at the census year, the latest of which is 2011, updated to 2019 by proportional factoring. As all this data is grounded at the place of residence, we need to work backward by disaggregating trips by occupation at the residence and then computing occupational classes at the workplace end of the trip. This then enables us to calculate the number of essential workers at both ends of the trip.

To indicate how we generate this data, trips between a workplace $i$ and a residential location $j$ are first defined by mode of travel $k$ as $T_{ij}^k$ from the updated Census data. This is consistent with employment $E_i$ in workplace $i$ and working population $P_j$ at residential location $j$ defined from $E_i = \sum_j \sum_k T_{ij}^k$ and $P_j = \sum_i \sum_k T_{ij}^k$. As we know the proportions of occupations $o$ at the residential end of the trip $\rho_j^o$, we can apply these first to generate a disaggregate pattern of trips $T_{ij}^{ko} = \rho_j^o T_{ij}^k$ from which we can compute both the working population and the employment by occupational group at the residential and workplace ends of the trip $P_j^o = \rho_j^o P_j = \rho_j^o \sum_i \sum_k T_{ij}^k$ and $E_i^o = \sum_j \sum_k \rho_j^o T_{ij}^k$. We also need to divide employment, working population and trips into essential ($es$) and non-essential ($ne$) workers. We have this for the residential end of the trip from more aggregate data where we define $\rho_j^o = \rho_j^o(es) + \rho_j^o(ne)$ and we then use these proportions to generate the essential and non-essential workers and working populations as

$$\left. \begin{array}{c} E_i^o(es) = \sum_j \sum_k \rho_j^o(es) T_{ij}^k \, , \quad E_i^o(ne) = \sum_j \sum_k \rho_j^o(ne) T_{ij}^k \\ P_j^o(es) = \rho_j^o(es) \sum_i \sum_k T_{ij}^k \, , \quad P_j^o(ne) = \rho_j^o(ne) \sum_i \sum_k T_{ij}^k \\ E_i^o = E_i^o(es) + E_i^o(ne), P_j^o = P_j^o(es) + P_j^o(ne) \end{array} \right\} . \quad (1)$$

We now need to consider the volumes of essential workers from the data before we embark on our first foray into examining the distribution of those who are still working during the first Lockdown. To get some sense of the variation in the distribution of essential and non-essential workers, we can aggregate the employment and working populations given in equation (1) to the UK and then to Greater London. In Table we show $E^o = \sum_i E_i^o$ and $E^o(es) = \sum_i E_i^o(es)$ and $E_{GLA}^o = \sum_{i \in GLA} E_i^o$ and $E_{GLA}^o(es) = \sum_{i \in GLA} E_i^o(es)$ and it is clear that there are substantial differences between different locations as well as between the volume of different occupations for different levels of spatial aggregation.

The way the government defined essential workers was based on particular subcategories of occupation whose families required support such as child care (Cabinet Office, 2020), and these subcategories varied in size for each occupational class in different areas and were defined from the Standard Occupational Classification (SOC) which has a very detailed breakdown of the 9 basic occupations. The percentages of workers in each basic occupation determined to be essential are illustrated in Table 1 for the UK and for London. It is immediately clear that the proportion of essential workers in London is much smaller than in the whole UK, 16% compared to 24% and although we do not have space here to examine these variations over the whole country, they are significant and thus make a big difference to the sheer volume of mobile workers during the Lockdown for the entire country. In fact in this project we have extended our analysis to England, Scotland and Wales using the Census geography of the middle-layer super output areas (MSOAs) of which there are 8436 in Great Britain (where we use the term Britain as a short hand for the three countries involved) but here we will focus only on London. Our figure of 24% working in essential services in the UKL compares well with the figure of 22% from the Institute of Fiscal Studies which is also based on the list of key occupations provided by government (Farquharson, Rasul and Sibieta, 2020) but applied slightly differently.

In Table 1, where we show total and essential workers by occupational categories for the UK and London, we first note that the relative importance of managerial and professional groups (the first three categories) which are much greater for essential workers in the UK than in London. The table also shows that the proportions of carers acting as essential workers in the UK is twice that of London. If we look at the last two columns in Table 1 where we have computed the proportions of non-essential and essential occupational groups for the UK and London in terms of the total employment of each area, we see that proportions of non-essential and essential managerial-professional groups is a little higher in London with skilled trades a little lower and caring and leisure services somewhat higher. A closer analysis of Table 1 suggests that London is weighted more to managerial occupations and service occupations than the UK in general but that in these occupations there are less proportions in the essential workers category than the UK in general. This bears out in very broad terms the fact that there are less non-essential workers in London with the implication that is non-essential are likely to be working from home, mobility levels will be a lot less than other parts of the country.

Our preliminary analysis of this data involves examining the relative distributions of essential and non-essential workers at their place of work and at their place of residence. In the analysis, we aggregated the occupational data, thus, working with essential and non-essential workers at their workplace and residence. These are defined from the above employments as

$$\left. \begin{array}{l} E_i(es) = \sum_o E_i^o(es) \\ E_i(ne) = \sum_o E_i^o(ne) \\ P_j(es) = \sum_o P_j^o(es) \\ P_j(ne) = \sum_o P_j^o(ne) \end{array} \right\} \quad . \qquad (2)$$

We initially speculated that essential workers would travel less distances to work than non-essential, especially in London and this would be reflected in their work and home locations. We would expect non-essential to be more clustered towards the centre of the city although the complications of the London housing market could well obscure such a clear pattern because central and inner London are now so highly priced. Therefor we might also be able to detect some evidence that essential workers might actually travel further to work so that they can access lower priced housing. The trade-off in London between house price and travel cost however is complicated and the data to measure this is problematic. If we first look at the correlations between essential and non-essential workers at their place of work for the UK aggregated now over occupations, this is 0.981 in comparison with their place of residence where it is 0.882. However the correlations between essential and non-essential between workplace and home are very low. In short, there is a very dramatic difference in the UK between where workers live and work which is perhaps somewhat greater than what one might have expected. Correlation is one of many measures we can use to look at this covariance and it does not tend to pick up the autocorrelation in these data but it does bear out the fact that there are big differences particularly in London where people live and work for both essential and non-essential workers. These bear out similar correlations to the UK.

*Table 1*. Total and Essential Workers by Occupational Categories for the UK and London
(Source: Cabinet Office, 2020)

| UK | Employment | Essential | Ess/Emp | NEss/Total | Ess/Total |
|---|---|---|---|---|---|
| 1 Managers, directors & senior officials; | 3,149,600 | 168,800 | 0.054 | 0.127 | 0.007 |
| 2 Professional | 5,532,200 | 1,170,300 | 0.212 | 0.186 | 0.050 |
| 3 Associate professional, technical occupations; | 3,854,300 | 919,600 | 0.239 | 0.125 | 0.039 |
| 4 Administrative and secretarial occupations; | 2,050,900 | 312,100 | 0.152 | 0.074 | 0.013 |
| 5 Skilled trades | 2,932,400 | 570,800 | 0.195 | 0.101 | 0.024 |
| 6 Caring, leisure & other service occupations; | 1,700,900 | 973,200 | 0.572 | 0.031 | 0.042 |
| 7 Sales & customer service occupations; | 719,700 | 71,800 | 0.100 | 0.028 | 0.003 |
| 8 Process, plant & machine operatives; | 1,754,200 | 796,200 | 0.454 | 0.041 | 0.034 |
| 9 Elementary | 1,752,400 | 542,500 | 0.310 | 0.052 | 0.023 |
| **Total** | **23,446,600** | **5,525,300** | **0.236** | **0.764** | **0.236** |

| LONDON | Employment | Essential | Ess/Emp | NEss/Total | Ess/Total |
|---|---|---|---|---|---|
| 1 Managers, directors & senior officials; | 617,300 | 14,800 | 0.024 | 0.130 | 0.003 |
| 2 Professional | 1,230,000 | 242,900 | 0.197 | 0.213 | 0.052 |
| 3 Associate professional, technical occupations; | 870,100 | 120,000 | 0.138 | 0.162 | 0.026 |
| 4 Administrative and secretarial occupations; | 412,100 | 52,100 | 0.126 | 0.078 | 0.011 |
| 5 Skilled trades | 317,800 | 46,000 | 0.145 | 0.059 | 0.010 |
| 6 Caring, leisure & other service occupations; | 328,200 | 80,200 | 0.244 | 0.054 | 0.017 |
| 7 Sales & customer service occupations; | 265,300 | 10,900 | 0.041 | 0.055 | 0.002 |
| 8 Process, plant & machine operatives; | 210,700 | 98,000 | 0.465 | 0.024 | 0.021 |
| 9 Elementary | 377,500 | 76,700 | 0.203 | 0.065 | 0.017 |
| **Total** | **4,629,000** | **741,600** | **0.160** | **0.840** | **0.160** |

Our data represent counts of workers in the small zones called MSOAs which on average have some 2,378 employees at their workplace and the same working populations at their residences. In fact the standard deviation for workplaces is much bigger than for residences (5,360 compared to 760) and this shows the very skewed distribution of the workplace data compared to the residential areas. To generate normalised distribution which are somewhat more comparable, we have defined densities as follows $E_i(es)/L_i$, $E_i(ne)/L_i$, $P_j(es)/L_j$, and $P_j(ne)/L_j$ where $L_i, L_j$ are the land areas of the zones in question. We have also correlated these variables for the UK and London and this shows a slightly different pattern; all these are shown in the heat maps in Figure 1 which, for counts and densities of the data in MSOAs, show almost exact correlations between essential and nonessential workers at their workplace and very strong correlations at their residence (home). Only for essential and non-essential workers at their residence is there any clear and obvious difference, and this is with respect to counts data where the absolute non-normalised size effect is more dominant.

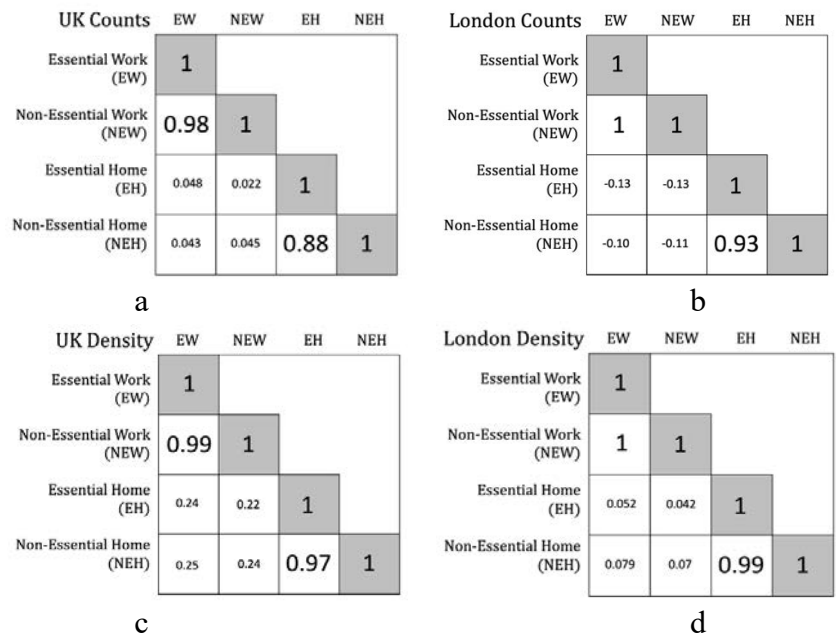

*Figure 1: Heat Maps Showing the Correlations for the UK and London*
For Count and Density Data a) UK Count Data, b) London Count Data, c) UK Density Data, and d) London Density Data

To complete this preliminary picture of how locations of essential and non-essential workers at their workplaces and residences are spatially distributed, we begin with the counts data that we show in Figure 2 for the UK and Figure 3 for London. Because these distributions are so skewed with very few large values and many small, we plot the logarithms of these data. The distinction between correlations associated with workplaces and residential areas in terms of essential and non-essential workers is reflected in these maps, first for the UK in Figure 2.

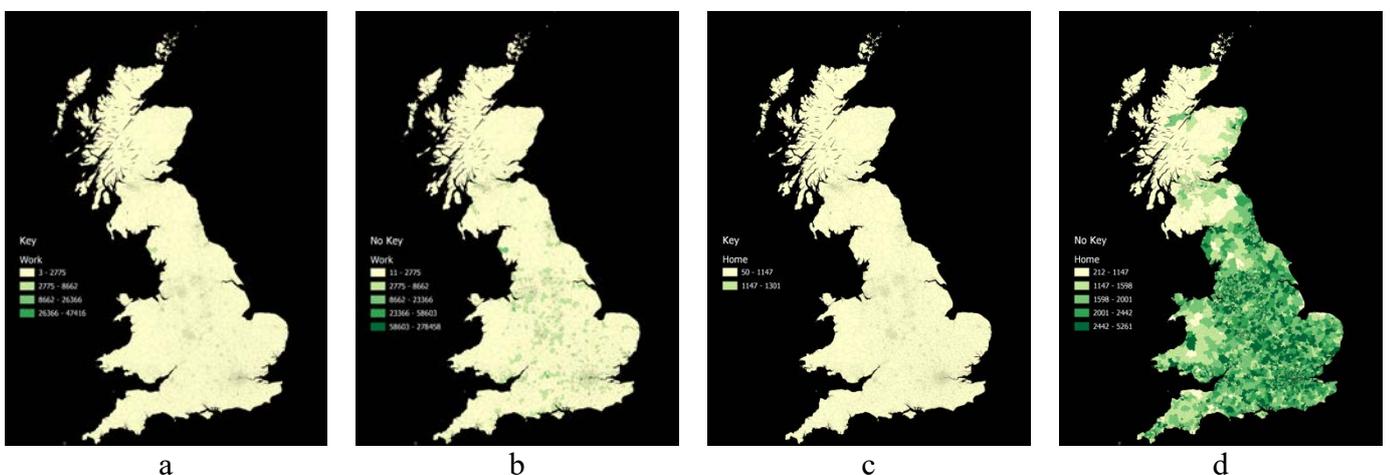

*Figure 2: UK: Essential and Non-Essential Employment at Workplace (a and b) and at Home (Residence c and d)*

It is clear that at the workplaces the UK distributions of essential and non-essential are very close as reflected in the correlations in the Heatmaps in Figure 1(a) and Figure 1(c). When we examine the home-based data in Figures 2(b) and 2(d), the biggest differences between essential and non-essential areas are in rural locations where travel times are much longer to reach workplaces and homes. Essential workers tend to be a little less concentrated near the bigger cities. We map the same distributions for London (the GLA area) in Figure 3 and the same sorts of conclusion emerge.

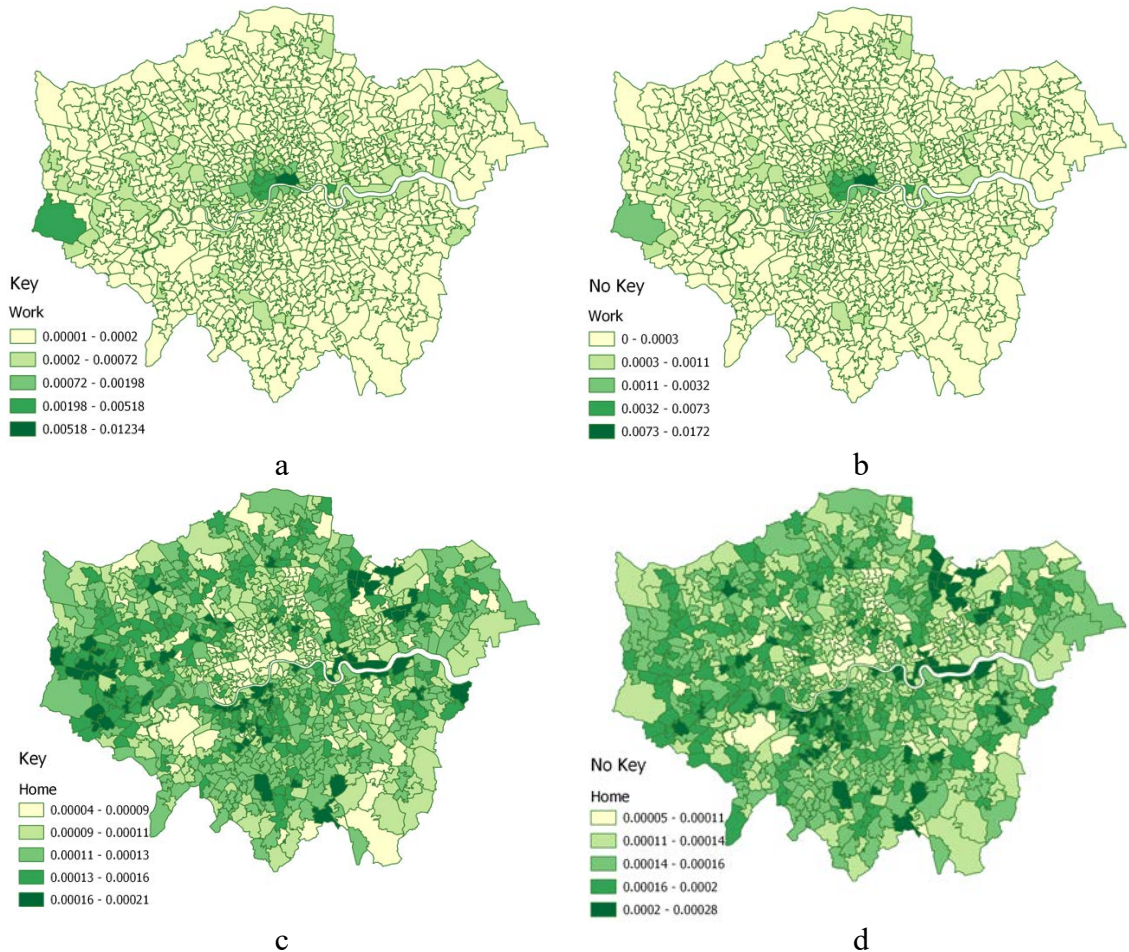

*Figure 3: London: Essential and Non-Essential Employment at Workplace (a and b) and at Home (Residence c and d)*

From Figure 3, the workplace distribution of essential workers is quite close to the non-essential for London and there is little to suggest any real spatial differences. In terms of the distribution of these same employments at the home-residence locations, the non-essential appear to be more clustered than the essential with the essential living a little closer to the centre. To explore these differences further, we can generate the same maps for the density distributions. As before, we first map the national density distributions of essential and non-essential workers at their workplace and home in Figure 4. From this first analysis, we can conclude that the spatial distributions of essential employment differs very little from non-essential at their place of work, as much because these distributions are very highly skewed and follow power laws. This is the reason we have been plotting them as logarithmic transforms to get rid of extreme variations and make them as comparable as possible, visually. At the home location, there is more variation but in terms of the UK and then London, it is more difficult to generalise other than saying that non-essential appear to be more randomly distributed than essential which are lower in rural and remote areas.

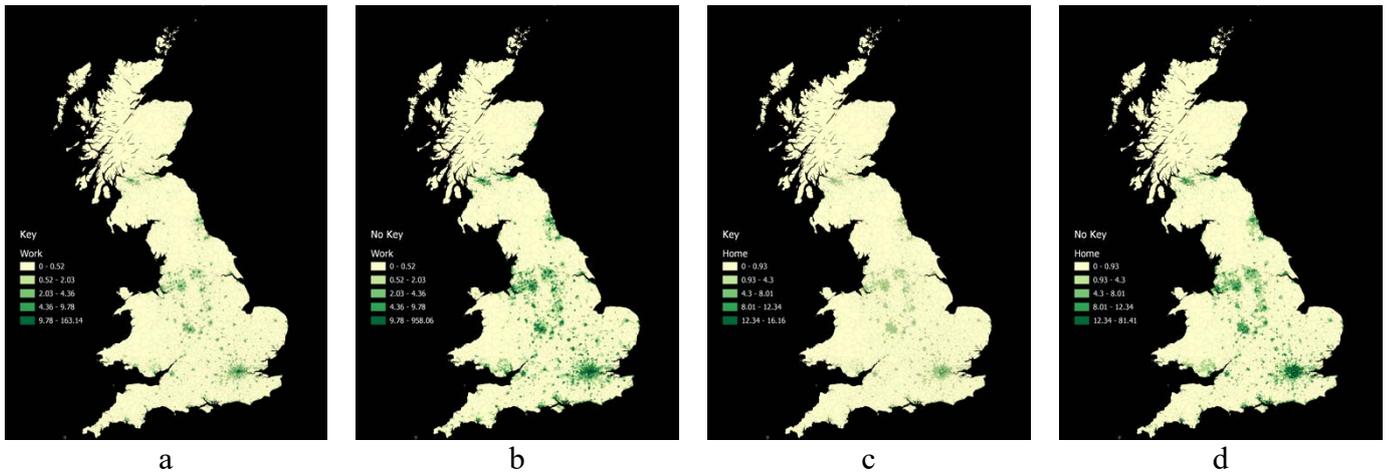

*Figure 4: UK: Essential and Non-Essential Employment Densities at Workplace (a and b) and at Home (Residence c and d)*

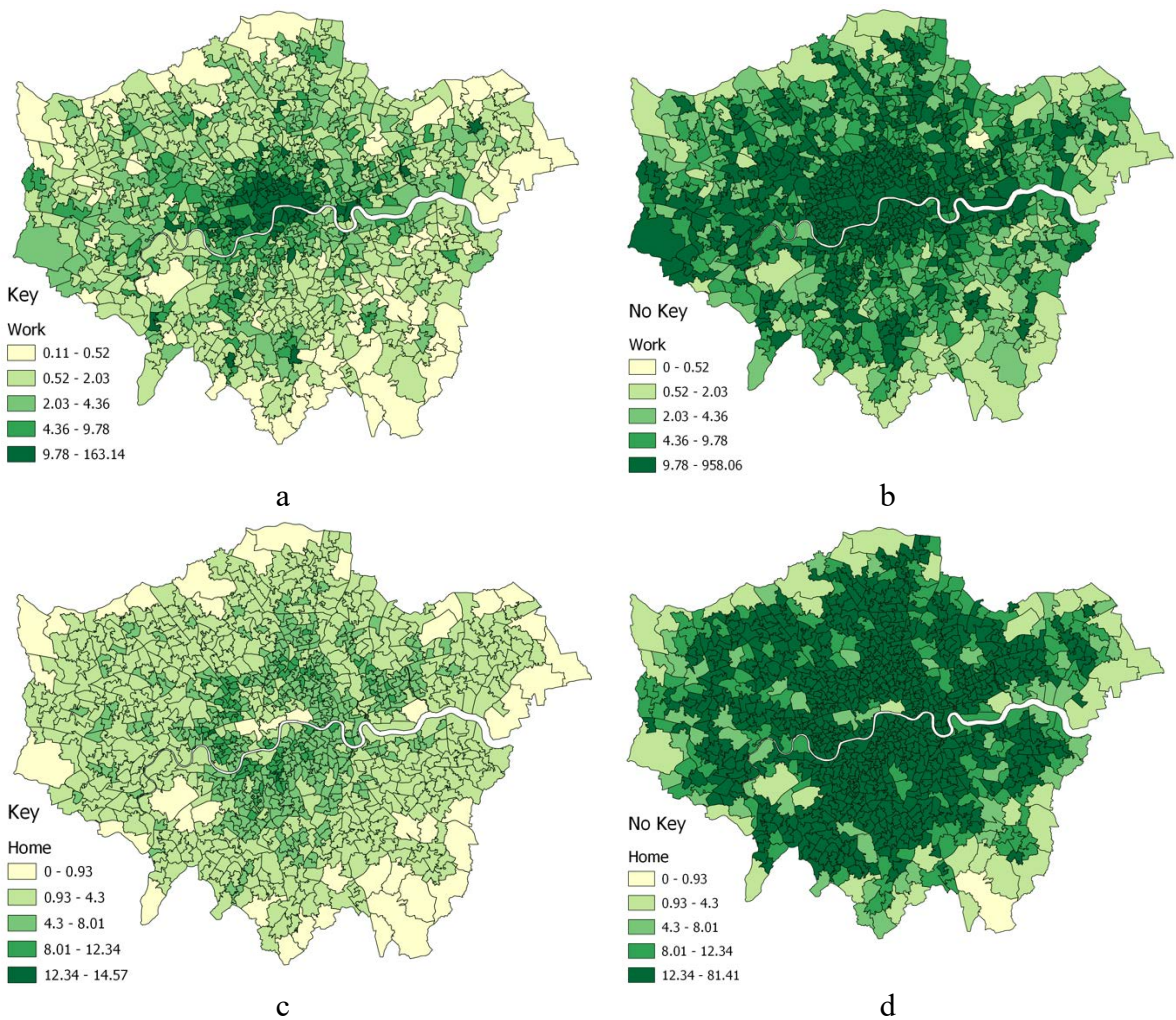

*Figure 5: London: Essential and Non-Essential Employment Densities at Workplace (a and b) and at Home (Residence c and d)*

## The Movement Patterns of Essential and Non-Essential Workers

Rather than simply focussing on how different are the location patterns for essential and non-essential workers, a more important aspect of the analysis is to examine the different trip lengths for the basic variable $T_{ij}^{ko}$. We do this for the overall distribution and then for the aggregations that we have introduced above. The mean trip length $C$ over the entire system measured in minutes is

$$C = \sum_i \sum_j \sum_k \sum_o T_{ij}^{ko} d_{ij}^k / \sum_i \sum_j \sum_k \sum_o T_{ij}^{ko} \qquad (3)$$

and then for the overall aggregations for occupations, modes and occupations by mode, these means are

$$\left.\begin{array}{l} C^o = \sum_i \sum_j \sum_k T_{ij}^{ko} d_{ij}^k / \sum_i \sum_j \sum_k T_{ij}^{ko} \\ C^k = \sum_i \sum_j \sum_o T_{ij}^{ko} d_{ij}^k / \sum_i \sum_j \sum_o T_{ij}^{ko} \\ C^{ko} = \sum_i \sum_j T_{ij}^{ko} d_{ij}^k / \sum_i \sum_j T_{ij}^{ko} \end{array}\right\} . \qquad (4)$$

We can also aggregate these to workplace zones and all sub-aggregations to occupations, modes and occupations by mode

$$\left.\begin{array}{l} C_i = \sum_j \sum_k \sum_o T_{ij}^{ko} d_{ij}^k / \sum_j \sum_k \sum_o T_{ij}^{ko} \\ C_i^o = \sum_j \sum_k T_{ij}^{ko} d_{ij}^k / \sum_j \sum_k T_{ij}^{ko} \\ C_i^k = \sum_j \sum_o T_{ij}^{ko} d_{ij}^k / \sum_j \sum_o T_{ij}^{ko} \\ C_i^{ko} = \sum_j T_{ij}^{ko} d_{ij}^k / \sum_j T_{ij}^{ko} \end{array}\right\} , \qquad (5)$$

and then to residential destinations

$$\left.\begin{array}{l} C_j = \sum_i \sum_k \sum_o T_{ij}^{ko} d_{ij}^k / \sum_i \sum_k \sum_o T_{ij}^{ko} \\ C_j^o = \sum_i \sum_k T_{ij}^{ko} d_{ij}^k / \sum_i \sum_k T_{ij}^{ko} \\ C_j^k = \sum_i \sum_o T_{ij}^{ko} d_{ij}^k / \sum_i \sum_o T_{ij}^{ko} \\ C_j^{ko} = \sum_i T_{ij}^{ko} d_{ij}^k / \sum_i T_{ij}^{ko} \end{array}\right\} . \qquad (6)$$

We need to note that the variable $d_{ij}^k$ is the travel time between origin $i$ and destination $j$ on the modal network $k$ of which there are three – bus, rail and road – but this does not include travel time at the trip ends. It is solely the travel time on the mode of transport – the time spent on the bus, in the train, or in the car. It is very likely that these times would double if the trip end times were added to them.

The breakdown of total modal trips into essential $T^k(es) = \sum_i \sum_j \sum_o T_{ij}^{ko}(es)$, non-essential $T^k(ne) = \sum_i \sum_j \sum_o T_{ij}^{ko}(ne)$, and total $T^k = \sum_i \sum_j \sum_o T_{ij}^{ko}$ is shown in Table 2 where it is clear that the percentage divisions for each mode do not differ very much from the 20-80 split which is roughly the overall split advised by government. In terms of the trip lengths, these are shown in Table 3 for each mode and overall and they differ very little between essential and non-essential. The minor differences in terms of proportions are such that what this probably means is that essential and non-essential trips patterns are very close to each other across all modes and this is echoed throughout this analysis. The aggregate mean trip length for the whole UK system $C$ is 15.89 minutes. When we disaggregate these into modes $C^k$, the longest is the rail at 30.63 minutes ($C^{k=2}$), followed by bus at 26.64 ($C^{k=1}$), and then there is a large drop to road (largely meaning car) with 12.46 minutes travelled on average ($C^{k=3}$). In terms of the modal split, the proportions in each mode are

$$\rho^k = E^k / \sum_k E^k = \sum_i \sum_j \sum_o T_{ij}^{ko} / \sum_i \sum_j \sum_o \sum_k T_{ij}^{ko} \qquad (7)$$

and these are calculated approximately as 10% for bus, 11% for rail and 79% for road.

*Table 2: UK Total, Essential and Non Essential Workers $T^k(es)$, $T^k(ne)$, $T^k$*

| | Total Workers | Essential Workers $T^k(es)$ | Non-Essential Workers $T^k(ne)$ | Total $T^k$ | % Essential | % Non-Essential |
|---|---|---|---|---|---|---|
| Road | | 3105661 | 12717370 | 15823031 | 0.20 | 0.80 |
| Rail | | 353080 | 1833292 | 2186372 | 0.16 | 0.84 |
| Bus | | 383304 | 1667827 | 2051131 | 0.19 | 0.81 |
| **Total** | | 3842045 | 16218489 | 20060534 | 0.19 | 0.81 |

Although we examine the trip lengths by mode in Table 2, we should also note the division of the country into its standard regions – namely Wales (W), Scotland(S) and 9 English regions – East Midlands (EM), East of England (EE), London (L), North east (NE), North West (NW), South East (SE), South West (SW), West Midlands (WM) and Yorkshire/Humberside (YH). The variations in trip lengths across modes still dominate the regional variations although there are some large deviations from the overall means. For example, in terms of road travel, the shortest travel times are in London (9.25) and the largest are in Wales (14.43) while rail travel is also smallest in London (22.98) and largest in Wales (45.82), the East Midlands (50.09) and the South West (64.88). The bus times are pretty even across all the regions with the largest being London but this is only 29.62 minutes compared to the average of 26.62, 11% more. Note that it is easy to confirm these statistics in that we can show if we add the mean trip lengths together for the three modes and weight them in the way we have shown in equation (7) above, then it is clear that

$$\sum_k \rho^k C^k = \left\{ \rho^1 \frac{\sum_i \sum_j \sum_o T_{ij}^{1o} d_{ij}^1}{\sum_i \sum_j \sum_o T_{ij}^{1o}} + \rho^2 \frac{\sum_i \sum_j \sum_o T_{ij}^{2o} d_{ij}^2}{\sum_i \sum_j \sum_o T_{ij}^{2o}} + \rho^3 \frac{\sum_i \sum_j \sum_o T_{ij}^{3o} d_{ij}^3}{\sum_i \sum_j \sum_o T_{ij}^{3o}} \right\} = C \ . \qquad (8)$$

A brief examination of the variances between these three modes between the 11 regions suggests that for bus, the standard deviation is about 6.68 minutes, followed by rail where it is 5.42, and then road which is 4.80. We can casually interpret these deviations as being due to the fact that most travellers have less control of the timing of their use of bus compared to rail and that the greatest control, hence the lowest variance, is for car use where the user has most control. However as all these effects are compounded across many zones and many travel times in different regions, it is not clear whether we can attribute such variation simply to the spatial differences across the nation or to the modes themselves.

When we examine the overall trip lengths by occupation and by region $C^o$, we find that managers and professionals travel some 33% more in time than the average while less professionally qualified occupations such as sales, technicians and some caring services travel some 25% less. London is a massive outlier where the managerial and professional occupations tend to have much less variation than in the regions but the less professional travel more than 40% of the national average. The North West and West Midlands tend to have lower travel times over most occupations than other regions. The singly-biggest difference with respect to occupations and regions is between an average travel time of some 20 minutes for professionals in all regions with the exceptions of the North West, West Midlands and Yorkshire-Humberside. The caring occupations only commute some 10-11 minutes while London is again the outlier and more peripheral regions such as Wales and Scotland do not appear to be dramatically different from the average. It is hard not to conclude from this brief analysis that London is dramatically different from the national average largely because of its size and the fact that its housing market and its transport systems are so different from the rest of the country.

We have one further disaggregation that we need to focus on and that is our basic distinction between essential and nonessential workers. In fact we can explore these through the occupations but as this would involve us in too much detailed analysis here, we will aggregate these essential and nonessential occupations into a total of essential and nonessential at the zonal, then the regional and the national levels. We will thus produce the same analysis that we have already developed for the essential and non-essential aggregates. Noting that we define the essential trips and non-essential as $T_{ij}^{ko}(es)$ and $T_{ij}^{ko}(ne)$, the mean trip lengths by mode as $C^k(es)$ and $C^k(ne)$, and by occupation as $C^o(es)$ and $C^o(ne)$, it is immediately apparent that the orders of magnitude of all these variables are very similar in values to $C^k$ and $C^o$. In fact, the essential and non-essential mean trip lengths for occupations and regions hardly reveal any differences from the combined distributions of all populations: the main difference is in the overall mean trip lengths $C(es)$ and $C(ne)$, which are 15.63 and 15.95, respectively, which is a difference of only about 2%. These are shown in Table 3 for the modes as well.

*Table 3: UK Mean Trip Lengths $C^k(es), C^k(ne), C^k$*

| Mean Costs | Essential Workers $C^k(es)$ | Non-Essential Workers $C^k(ne)$ | Total $C^k$ |
|---|---|---|---|
| Road | 12.52 | 12.44 | 12.46 |
| Rail | 31.23 | 30.52 | 30.63 |
| Bus | 26.48 | 26.68 | 26.64 |
| **Total** | 15.63 | 15.96 | 15.89 |

In short, this means that on average, essential workers only travel about 2% less than nonessential workers over all modes. The occupation data suggests all changes are less than 2% and these do not vary much within regions. In fact each of the 9 categories of worker by occupation has essential and non-essential workers and in general over all occupations, the essential travel is only very slightly less than the nonessential. In terms of modes, there is no more difference than between regions and occupations and it would appear that the essential workers are distributed in a very similar way to the nonessential across the country, probably due to the fact that to keep the system running, one has to have roughly the same pattern of workers everywhere. This is particularly pronounced for London as Table 4 reveals. In short the differences are hardly worthy of comment and our anticipation that essential workers differ in their journey patterns radically from non-essential is not borne out. In fact it is more likely that essential and non-essential differ in their spatial locations than in the amount of time spent in travelling but what we require is a new framework to handle all these variations so that we can apportion the relative importance of minor differences between category types.

Table 4: London Mean Trip Lengths $C_{GLA}^{k}(es), C_{GLA}^{k}(ne), C_{GLA}^{k}$

| Mean Costs | Essential Workers $C^k(es)$ | Non-Essential Workers $C^k(ne)$ | Total $C^k$ |
|---|---|---|---|
| Road | 9.12 | 9.28 | 9.25 |
| Rail | 22.98 | 22.98 | 22.98 |
| Bus | 29.76 | 29.59 | 29.62 |
| **Total** | 18.86 | 19.16 | 19.12 |

If we now look at the total amount of travel, rather than the trip lengths, we can begin noting that we can multiply the mean trip lengths by the total trips for whatever aggregation of the trips we are dealing with from $T_{ij}^{ko}(es)$ and $T_{ij}^{ko}(ne)$. First, we will look at total travel in the essential and non-essential sectors and these are defined as $T(es) = \sum_i \sum_j \sum_k \sum_o T_{ij}^{ko}(es) d_{ij}^k$ and $T(ne) = \sum_i \sum_j \sum_k \sum_o T_{ij}^{ko}(ne) d_{ij}^k$ which add to the total travel in the system as

$$T = \sum_i \sum_j \sum_k \sum_o T_{ij}^{ko} d_{ij}^k = \sum_i \sum_j \sum_k \sum_o [T_{ij}^{ko}(es) + T_{ij}^{ko}(ne)] d_{ij}^k = T(es) + T(ne) \quad (9)$$

These statistics suggest that before the Pandemic, the total amount of travel for work during the day was about 5.31 million hours per day in the UK as shown in Table 5. If we look at essential workers, the number of hours travelled after the lock down was about 1 million hours which means that some 4.31 million hours has been saved by persons working from home. This is a fall of some 81% in travel time which reflects the number of essential workers in the whole economy. It is worth saying that this is not the total reduction in travel because people working from home still go shopping and take exercise. Note that if we divide these total travel times by the respective total trips, then this gives the mean travel times, in this case of $C(es) = 15.63$, $C(ne) = 15.95$, and $C = 15.89$.

Table 5: UK Total Hours Spent in Essential Travel and Saved in Home Working $C^k T^k / 60$

| Hours | Essential Workers | Non-Essential Workers | Total |
|---|---|---|---|
| Road | 648,048 | 2,636,735 | 3,285,916 |
| Rail | 183,778 | 932,535 | 1,116,143 |
| Bus | 169,165 | 741,627 | 910,702 |
| **Total** | 1,000,853 | 4,314,118 | 5,312,698 |

*Drilling Down Into Individual Locations in London*

So far, we have only examined the aggregate pattern of essential and non-essential workers at their workplace and home where the distribution of workers has been mapped prior to the Pandemic. Then those who are furloughed and work from home who are deemed non-essential are subtracted from the total leaving those who are essential workers dominating the journey to work during the 3 months period from March 23rd 2020 when the country was first locked down. However, we are able to approach the problem of the fall in travel in a more oblique manner using Google Mobility Reports (2020) which collate data from anyone who users

Google services on mobile devices such as Google Maps and who has switched on their Location History (which is off by default). This is available for many places around the world and in the UK is available for all local authorities. In London, this means that we have good data on the changes in trips day by day from 10 February 2020 to 18 October 2020 for six categories of activities: residential occupancy, workplace volumes, visits to parks, volumes of workers using transit which in the case of London is overground and underground rail and buses, retail sales, and then grocery store volumes of visits. What these data show is the decrease or increase in volume from the baseline. In this case for the UK, it is some 37 days before Lockdown and covers the period from then to our cut-off date when we retrieved the data giving us a series of 247 days. The horizontal axes of the trajectories that we show below in Figures 6 to 9 span these 247 days on a scale from 0 to 250. In fact, the baseline is the median value of activity volumes in the 5-week period from 03 January to 06 February 2020. The data continues to be collected.

For each place and aggregation thereof, the data provides a detailed time series of the level of activity before the Lockdown with the baseline data as 10 February. Then the subsequent change in activity which at the beginning of the period is a massive drop in visits due to the fact that people were mandated not to travel (as we noted in the introduction) suggest a slow recovery towards the baseline. In the case of the UK, and London in particular, there is no return to the baseline and there has in fact been a levelling off and even a slight drop in some activities in the last 40 or so days as a second wave of infection appears to be beginning. There is a wealth of data here but with all the caveats of course pertaining to its representativeness for only if you have Location History on and access to Google services can any such data be collected. The representativeness of the data is thus unknown and only if the data were integrated with other sources would we be able to say very much about its overall quality. What we can say, however, is that the data is accurate in terms of its geo-positioning.

Our analysis in this paper merely touches the surface of what we can do with data such as this and as yet we have not embarked on a full scale analysis of all areas of the UK and any aggregation thereof. What we do here is begin with the data for the UK and use this as our spatial baseline, thence comparing three different parts of the Greater London Authority with this baseline and with each other. These locations are the City of London – the so-called dead heart of the metropolis (Gruen, 1964), the more prosperous borough of Richmond-on-Thames in west London and the less prosperous borough of Newham in east London. The average income/wage of those in the City is about £27 per hour (ph), in Richmond it is £20 ph and in Newham £13 ph, all compared with a London average of £15 ph and a UK average of £13 ph. This gives a very crude idea of the relative prosperity of these places and we might expect this to have some effect on the degree to which mobility patterns might have altered during the Pandemic. There are many other estimates of income but these appear to be the least controversial (see London Data Store, 2020; and ONS, 2020).

We show the profiles of the change in the six activities for the UK in Figure 6 and the same for the three London boroughs – the City in Figure 7, Richmond on Thames in Figure 8, and Newham in Figure 9. For the UK, the growth in those working at home peaked early in the Pandemic at about an additional 30 percent staying home to work or furlough, then gradually falling back to about 10 percent more than the baseline. There is no data for the City of London as only 8,000 people live there and many only do so during the working week but the richer borough of Richmond had up to 50 percent and this has fallen back to some 25 percent now. Newham is similar to the UK baseline. The only other activity which has seen any real increases in activity are in parks, which fell nationally in the early and then quickly increase to a peak of 150 percent about average through the summer months well above the baseline. This has fallen back to some 25 percent above the baseline. Richmond is similar to this profile but in Newham there has been massive increase up to 500 precent of above the baseline of people visiting parts where in the City of London where half a million workers use pocket parks during the day, volumes have fallen by over 90 percent rising back to some 60 percent below the baseline but now back down to 70 percent as winter approaches.

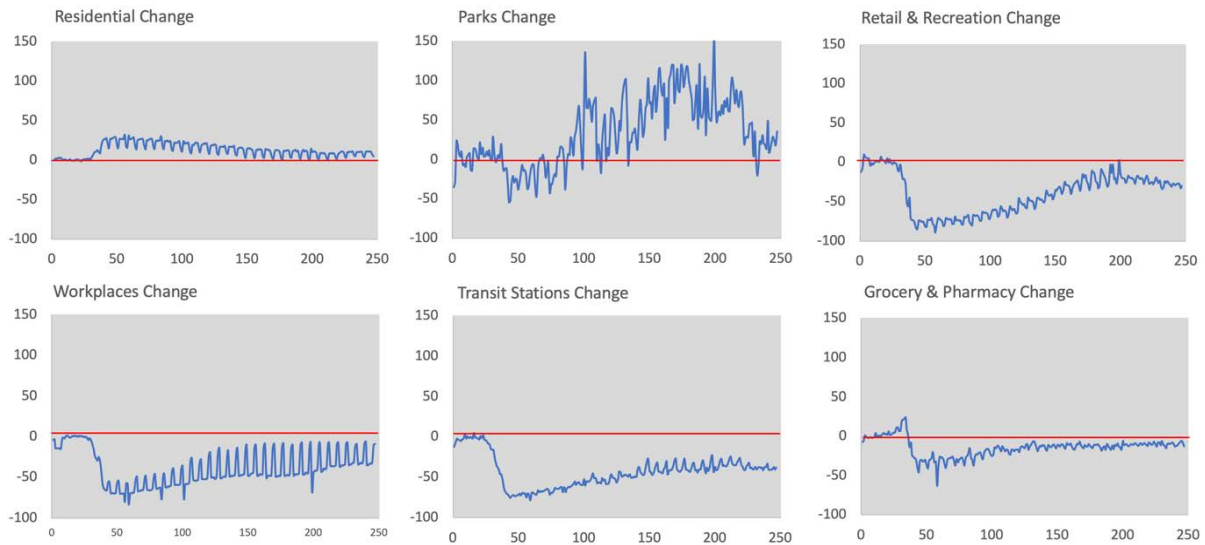

*Figure 6: National UK Percentage Changes in Use of Activities from the February Baseline*

The horizontal axes relate to the number of days from the 10th February and the vertical axes are the negative or positive percentage shifts in activity volumes from the baseline. This applies to all subsequent figures.

The other four activities – retail, workplaces, use of transit and grocery shopping – all show a big drop when Lockdown took place and then a gradual recovery but only in the case of grocery shopping has this recovered almost to the baseline – the old normal. This is apart from the City where grocery shopping is mainly based on office workers who are no longer at work. Retail change is startling with a national fall by about 90 percent at the beginning of the and then a steady rise back to about 20 percent below the baseline by the late summer but with a distinct fall back to about 35 precent below the line during the last 50 days. There is evidence here of a second peak but this is compounded by a drop in income and as yet the picture is unclear. Richmond and Newham are fairly similar to the national UK profile apart from two massive spikes in Richmond on 23rd February and 7th March which relate to key Rugby matches at Twickenham where England played Ireland and Wales in the 6 Nations Cup in the days before the Lockdown occurred. The same picture occurs in transit usage with spikes in Richmond but quite dramatic falls in usage in all four examples. In the UK, volumes using transit fell to about 70 percent below the baseline but then have recovered to about 50 percent, while in the City these have fallen even more by about 95 percent but have only recovered to within some 70 percent of their normal baseline. Richmond and Newham have a similar profile to the UK but have only recovered to about 60 percent of their previous normal volumes. This is a major problem in that public transport is being completely subsidised by government and local authorities throughout England and is effectively bankrupt. National Rail have more or less been re-nationalised and Transport for London is in negotiation with the UK government. Congestion charging has been increased already in London and fare rises much greater than inflation have been suggested but not yet agreed. The public transport picture is much confused and at the time of writing, where a new national Lockdown is under discussion, the picture for transit looks bleak.

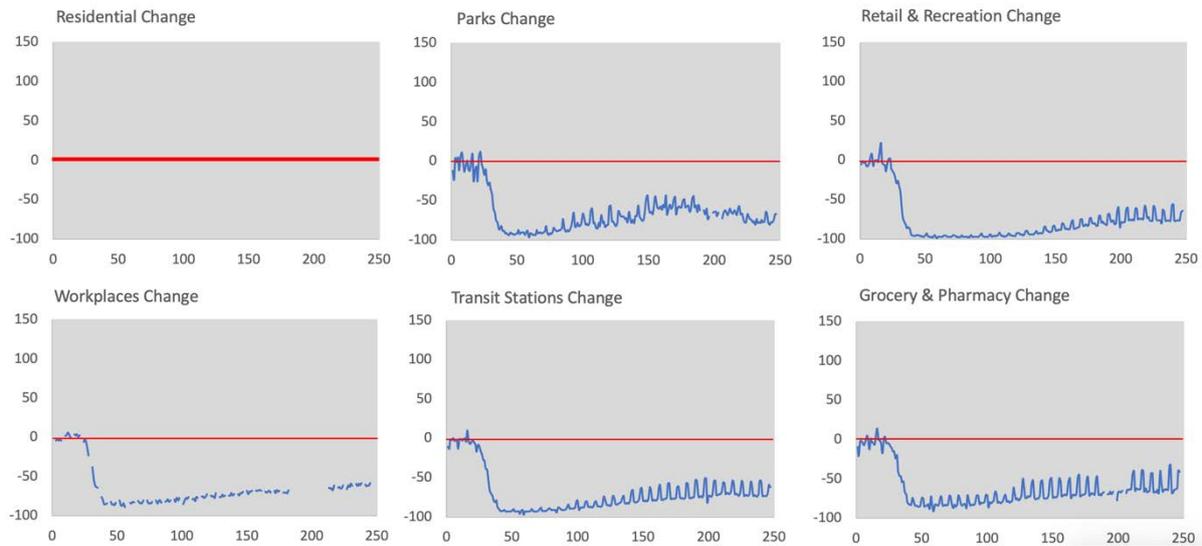

*Figure 7: The City of London Percentage Changes in Use of Activities from the February Baseline*

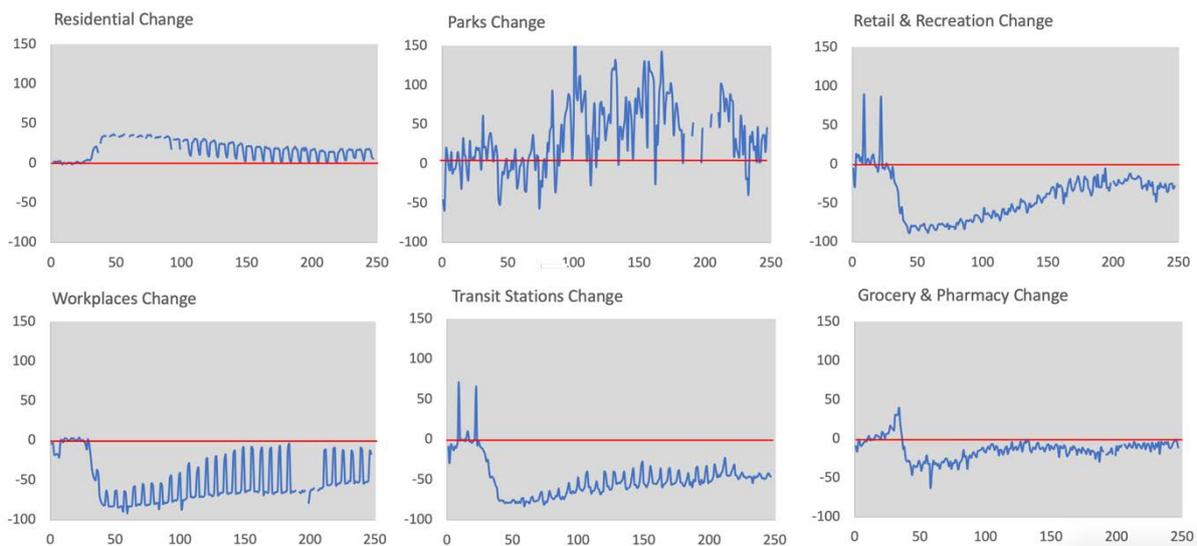

*Figure 8: The Borough of Richmond on Thames Percentage Changes in Use of Activities from the February Baseline*

The key changes of course with respect to the focus here on essential and non-essential workers pertain to change in workplace volumes which imply changes to those working in their traditional workplace and those who have elected to work from home. When we look at the national picture in Figure 6 for workplace change, the profile is fairly typical of other declines: the national fall is about 70 precent which then recovers to some 30 percent less than the baseline. In Richmond, the decline is similar to the UK and to Newham but the decline in the City of London is much more fractured due to gaps in data collection during the weekends while in the three other cases the impact of the weekend versus the weekday is very pronounced. In short, the succession of spikes along the trend from a big fall and then recovery back towards the baseline marks the way the weekend and weekdays impact on work journeys to places of work. Without more information, it is difficult to precisely gauge these movements but these are quite pronounced in Richmond where the impact of the weekday seems to generate many less workers than the weekends.

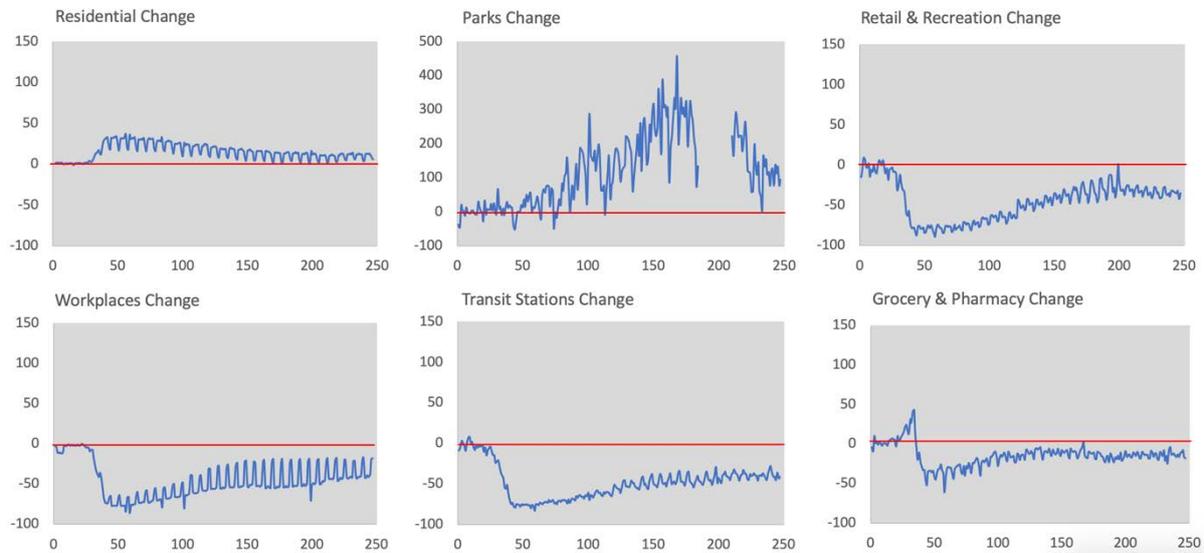

*Figure 9: The Borough of Newham Percentage Changes in Use of Activities from the February Baseline*

The last activity we need to examine is very local – grocery shopping which drops the least of any activity after Lockdown – to no more than 50 percent of normal activity. In all but the City, this recovers to near normality by the end of the period. In fact, there is spike at the onset of the Pandemic with grocery shopping increasingly above the baseline, largely we suspect when people were stocking up on products ready to shop far less in the early months of the Pandemic. This is the case for three of our cases but not the City of London where there is a pronounced drop to about 90 percent of normal activity which slowly climbs back but to no more than 50 percent less than the baseline. This is largely due to the fact that most grocery shopping in the City is by workers from their place of work, not their home and it is interesting that these features are reflected in the data. This is also true of the spikiness of the trajectories which reflect differences between home and work with the distortions posed by working from home, affecting this drop in activity much more on weekdays than is represented on weekend days.

Overall, these trajectories confirm what we know from our casual experience of the Pandemic, but what is interesting is that they do give us some insight into changes in activities even though they only come from one source, Google. We cannot prove that this data is representative but little differences such as the impact of sporting events which show up in the data are likely to show up in similar activity patterns from other mobile operators. Moreover, what is of current interest is in how these patterns will be reflected in another Lockdown. As we have had a series of partial Lockdowns of varying severity in the UK during the last 3 months, it may be possible to trace the impact of these once the full Lockdown becomes active on 05 November. Although this data is useful, it is hard to tie it to independent data sets, and during the Pandemic it is difficult to mount special surveys to examine more conventional flow data, notwithstanding the work that the Office of National Statistics are undertaking.

## *Conclusions and Next Steps: A More Integrated Analysis*

The most surprising conclusion from the analysis involves the dramatic differences between the regions of the UK with respect to average travel times on different modes of transport, and the minimal differences between average travel times between essential and non-essential workers in each region. To an extent, this may be due to the fact that the way the government has defined essential workers is close to the distribution of non-essential workers and if this is the case, the transport networks and modal split will not make any real difference to the patterns of distribution which essentially are the same for these two groups.

A key issue which is raised by this paper is data. From data prior to the Pandemic, which is the only detailed data we have on flow patterns and locational volumes, it is of little surprise that we cannot generate many new insights into spatial differences that might be exploited in handling the Pandemic. To this end we need much better data on the spatial progression of the Pandemic, and we need to tie the sort of data produced by Google (with equivalent mobility data from IT platforms such as Apple (2020) from their map products, and Amazon Web Services, etc.) to more conventional transport data such as traffic counts from sensors, data from smart card ticking systems and so on.

What we need to do now is to regionalise the country into localities and regions that might be highly clustered with respect to similarities (and differences in an alternative analysis) so that we can examine where there are significant differences that might be exploited in countering the Pandemic. In part of the country where people travel more or travel less than the average, this has implications for how we might lockdown certain areas. Currently, many governments in Western Europe are moving towards a total Lockdown to combat the second wave but these kinds of blunt instrument approaches are too crude to effectively exploit the differences in the way the wave of infections is diffusing. In this work, what we urgently need, as in all work on the spatial, is some way of linking the location of the incidence and intensity of the disease to human behaviours pertaining to location and mobility. This is essential and probably requires moving to a much more individual scale where we can tie health data to the location of workers and the population. This will require us to disaggregate our data down to the individual and household level so that we can then associate this with the diseases and susceptibility to disease, before we then aggregate this back up so we can link these attributes to those we have explored in this paper, which focus on how people move and how movement is connected to location.